\documentclass[useAMS]{mn2e}
\usepackage{url}
\usepackage{psfig}
\usepackage{color}

\def\etal{{\rm et~al.\ }}
\def\hmpc{\;h^{-1}{\rm Mpc}}

\def\hkpc{h^{-1}{\rm kpc}}

\def\msun{{h^{-1} M_{\odot}}}

\def\simlt{\lower.5ex\hbox{$\; \buildrel < \over \sim \;$}}
\def\simgt{\lower.5ex\hbox{$\; \buildrel > \over \sim \;$}}

\title[Halo occupation]{
Dark matter halo occupation: environment and clustering
}

\author[R.A.C. Croft \etal]{\parbox{18cm}{
Rupert A.C. Croft$^{1,2}$\thanks{E-mail: rcroft@cmu.edu},
Tiziana Di Matteo$^{1,2}$,
Nishikanta Khandai$^{1,2}$,
Volker Springel$^{3,4}$,
Anirban Jana$^5$,
and Jeffrey P. Gardner$^6$}\vspace{0.3cm}\\
$^{1}$ Bruce and Astrid McWilliams Center for Cosmology,
Carnegie   Mellon  University,\\
$^{2}$ Dept.   of  Physics,   Carnegie   Mellon  University,
Pittsburgh, PA 15213, USA\\
 $^3$Heidelberger Institut f\"{u}r Theoretische Studien,
        Schloss-Wolfsbrunnenweg 35, 69118 Heidelberg, Germany\\
   $^4$Zentrum f\"ur Astronomie der Universit\"at Heidelberg, Astronomisches
        Recheninstitut, M\"{o}nchhofstr. 12-14, 69120 Heidelberg, Germany\\
$^5$ {Pittsburgh Supercomputing Center, 300 S. Craig Street, Pittsburgh, PA 1521
3, USA} \\
$^6$ {University of Washington, Department of Physics, Seattle, WA 98195-1560, U
SA}
}
\begin{document}
\pagerange{\pageref{firstpage}--\pageref{lastpage}} \pubyear{2005}

\maketitle

\label{firstpage}

\begin{abstract}
We use a large dark matter simulation of a $\Lambda$CDM model to
investigate the clustering and environmental dependence of the number of
substructures in a halo. Focusing on
redshift $z=1$, we find that the halo occupation
distribution is sensitive at the tens of percent level
to the surrounding density and to a lesser extent to
asymmetry of the surrounding density distrbution.
We compute the autocorrelation function of halos
as a function of occupation, building on the finding of
Wechsler \etal (2006) and Gao \& White (2007) 
that halos (at fixed mass) with more substructure are more clustered.
We compute the relative
bias as a function of occupation number at fixed mass,
finding a strong relationship.
At fixed mass, halos in the top $5\%$ of occupation can have an autocorrelation 
function $\sim 1.5-2$ times higher than the mean.
We also compute the bias as a function of halo mass, for fixed halo
occupation. We find that for group and cluster sized halos, when
the number of subhalos is held fixed, there is
a strong anticorrelation between bias and halo mass.
Such a relationship represents 
an additional challenge to the halo model.
\end{abstract}


\section{Introduction}

The halo model of galaxy clustering (see Cooray and Sheth 2002 for a review)
 takes as
one of its usual assumptions that the clustering of a dark matter
halo is dictated solely by its mass.
Observationally when looking at a cluster or group it is
easiest to count galaxies (see e.g.,  Hao \etal 2010, Yang \etal 2005,
for recent
cluster and group catalogues),
which are usually associated with the
population of dark matter subhalos. The role that the number of
subhalos plays in the clustering of halos and also how that number is
affected by environment has the potential to be linked directly with
observations. If halo mass and number of subhalos affect clustering
independently then this is a challenge to a main
assumption of the halo model. In this paper we examine
the role of the number of subhalos in a halo on clustering in 
two different ways, first by studing the environments (nearby overdensity)
of halos and second by computing the correlation function.

Gao, Springel and White (2005) showed that the clustering strength of
dark matter halos depends not only on mass but also on formation time. This
result used a high resolution large volume simulation (the Millennium Simulation
of Springel \etal 2005) and was the first of a set of results
showing that halo mass is only the first order driver of halo clustering.
Subsequent studies confirmed the result (e.g., Zhu \etal 2006,
 Harker \etal 2006) and showed that
other dependencies exist apart from halo mass, such as concentration
(e.g., Wechsler \etal 2006, hereafter W06),
and halo spin and shape (Bett \etal 2007).
W06 also found a dependence on number of subhalos, and Gao 
and White (2007) (hereafter GW07) on the substructure fraction, both 
at fixed mass
in that halos with more substructure are more clustered.
 It is  these relationships that we will be investigating in more 
detail in this paper.

The clustering of halos can be probed in a different, related fashion
by measuring properties of the local environment. Again in this
case to first order the dependence of halo properties  on environment
was shown to be limited to mass by Lemson \& Kaufmann (1999). 
Advances in simulation size and resolution enabled this type of analysis to
be extended to measure smaller effects and different statistics. For  
example, the abundance of halo substructure in a high resolution
simulation of side length $21.4 \hmpc$ was shown by Ishiyama \etal (2008)
to be dependent on local overdensity. Wang \etal (2011) find that the
large-scale tidal field significantly affects 
all halo properties they studied (assembly time, spin,
shape and substructure) at fixed mass. White \etal (2010), concentrating
on cluster-sized halos show that their environment influences and
causes physical correlations in many observational probes of mass
and richness (e.g., subhalo
number, lensing and velocity dispersion).

The analysis in this paper is an extension of some of this prior work, which 
has already shown in many ways that the subtleties of halo clustering
are not explained by the simplest forms of the halo model. Here we
use the language of the Halo Occupation Distribution 
(HOD, Berlind \& Weinberg 2002) to
examine the environmental dependence of substructure.
In looking at clustering, we examine samples of halos from the point of view
of their bivariate distribution (mass and number of subhalos), which it
turns out can lead to some interesting behaviours which may suggest 
further improvements in the halo model.

The structure of the paper is as follows. 
In Section 2 we briefly describe the N-body simulation we use to 
study the subhalo population of halos along with their large-scale
enviroment. We give our measures of environment and describe our method
for computing statistics that depend on halo occupation at fixed halo mass.
The environmental dependence of the HOD is examined
in Section 3 and a resolution test is carried out. In Section 4 
we examine the clustering of halos, measuring their large-scale bias 
as a function of mass and subhalo occupation. In Section 5 we summarize
our results and discuss their implications for both the theoretical
halo model and observational probes of dark matter halos.

\section{Simulation}
We have used P-{\small GADGET}, an upgraded version of 
{\small GADGET}3
(see Springel 2005 for details of an earlier code version) 
which is being developed for upcoming
petascale supercomputer facilities, to run a large dark matter
simulation of a $\Lambda$CDM cosmology. The cosmological parameters
used were: amplitude of mass fluctuations, $\sigma_{8} = 0.8$,
spectra index, $n_{s} = 0.96$,  cosmological constant
parameter $\Omega_{\Lambda} = 0.74$, and 
mass density parameter $\Omega_{m} = 0.26$.
 In the present work we use the simulation  output
at $z=1$, the same as that analyzed in our previous paper using that 
simulation (Khandai \etal 2011).
The initial conditions were generated with the Eisenstein and Hu (1998)
power spectrum at an initial redshift of z = 159.  The
basic simulation parameters are: box side length, 400 $\hmpc$,
the number
of particles, $2448^3=1.5\times10^{10}$,
the mass of a  particle, $3.1\times10^{8} \msun$  and
the gravitational softening length, $6.5 \hkpc$.
For reference, our simulation volume
is roughly half that of the Millennium Simulation (Springel \etal
2005) but our mass resolution is about a factor of three better.

\subsection{Halos and Subhalos}
Halos and subhalos are identified on the fly as the 
simulation runs, the halos using a standard 
Friends-of-Friends (FOF) groupfinder with
linking length 0.2 times the mean interparticle separation.
The halo masses quoted are the sum of masses of all particles in the FOF group.

We use the {\small SUBFIND} code (Springel et al. 2001) to construct a
subhalo catalogue and to measure the mass 
for every subhalo. Groups of particles are
retained as a subhalo when they have at least 20 bound particles,
which corresponds to a minimum group mass of $M_{\rm sub}$ =
$6.3\times10^{9} \msun$. We will investigate the dependence of our results
on the minimum cutoff mass in the halo and subhalo masses. We also
carry out a resolution test in Section 3.2

The number of FOF halos in our simulation with mass greater than
$10^{10}\msun$ is 15.3 million. These FOF halos contain a total of
18.6 million subhalos (we do not differentiate between central and satellite
subhalos). Several of the trends we will be looking at are relatively
subtle and the statistical power coming from the large number of subhalos
and halos is needed to make them readily detectable.

We note that other authors have investigated the dependence
of substructure on other halo properties (e.g., 
Jeeson-Daniel \etal 2011, Skibba \& Maccio 2011) and clustering
on substructure (e.g., GW07, Wang \etal 2011). Substructure can be defined
in many different ways (such as the mass fraction in subhalos above a certain
mass) and also be counted either within the FOF halo or within a certain
radius (GW07 try $r_{200}$). We use the definition
of substructure appropriate to the usual definition of the HOD, the number of
subhalos above a certain mass within the FOF halo. We expect
(as shown by GW07 and Bett \etal 2007) 
that the complex nature of the relationship
between detailed halo properties and clustering will mean that different
definitions of substructure can yield  different results.

\begin{figure*}
\centerline{
\psfig{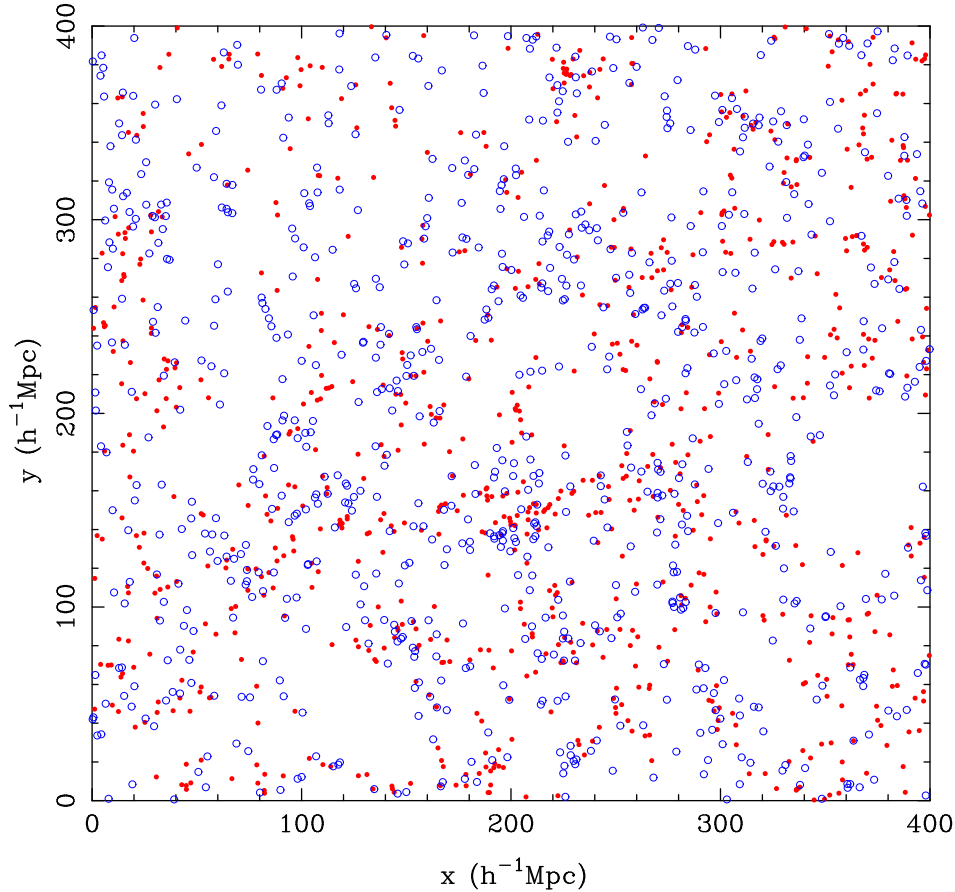}
}
\caption{
The locations of the top $5\%$ of halos by occupation at fixed halo mass
(see Section 2.3)
(red points) and bottom $5 \%$ (blue circles). We show
only halos with a mass $>10^{12}\msun$ in a slice of
thickness $40 \hmpc$.
\label{slice}
}
\end{figure*}

\subsection{Environment}
 As a measure of environment we have a  number of choices. The most
standard is the overdensity within a sphere of fixed radius.
We take this, using a fixed radius
of $5 \hmpc$ comoving to be our fiducial measure. We also test other 
radii, as well as 
 using instead the overdensity in the sphere centered
on a halo that encloses a fixed mass $10^{15}\msun$.
This Lagrangian definition of
overdensity has been used before in the context
of simulations by for example Colberg \& Di Matteo (2008), who investigated
the relationship between supermassive blackholes and their
environments. 

We note that the overdensity within either a 
fixed radius or Lagrangian volume is expected to correlate strongly
with halo mass (see for example the extensive study carried out
by Haas \etal 2011). Haas \etal 2011 show that it is possible to 
formulate measures of environment that are independent of halo mass.
In the present paper we have the more limited goal of investigating the 
dependence of number of subhalos (rather than mass) on environment. 
We can sidestep the mass-environment correlation by either plotting
a statistic we are interested in as a function of halo mass directly 
(the HOD) for different environments, or else
by plotting a statistic (the correlation function and its relative
bias) as a function of number of subhalos for fixed mass. Either way
we have removed the dependence on mass for the purposes of our
analysis.

\subsection{Halo occupation at fixed halo mass: definition}
Once we have a halo catalogue and have enumerated the subhalos per halo,
we would like to break up our full sample into high and low occupation
halos, where ``occupation'' is simply the number of subhalos (we
do not differentiate between central subhalos or satellites).

 In order 
to avoid dependence of properties (such as clustering) on halo mass,
we would like to somehow make the sample be effectively at fixed halo mass.
One possible approach involves applying
an upper and lower mass threshold to the halo sample
and then ordering the remaining halos by occupation. There are two disadvantages
with this. First, applying mass cuts in this way will 
reduce the number of halos available for study, perhaps dramatically if the
mass window allowed is narrow. Second, if the mass window is not
very narrow one could legitimately worry that dependence on halo mass will
still creep in, as there is obviously a strong dependence of halo
occupation and halo mass. This means that for such a mass window,
the top say 10\% of halos by occupation could be significantly 
more massive than the bottom 10\%. 

We avoid both of these problems by instead breaking our full halo sample
into a large number of narrow bins in mass. In each one of the mass bins,
we order the halos by occupation number per
unit mass, choosing the top 5\% of
halos, top 10 \% and so on. When making a sample of halos chosen
by occupation we then take the required fraction from each mass bin, so
that we are left with a sample that spans the entire mass range, but that 
was chosen by occupation at fixed halo mass. We have tried varying the 
width of the mass bins, $\Delta\log_{10}(M)$,
finding that below $\Delta\log_{10}(M)=0.5$ our results are independent
of its value (we use $\Delta\log_{10}(M)=0.2$). 

We make first use of our set of halos ranked by occupation at fixed
halo mass in Section 2.4 below when plotting their spatial distribution. The
main use for this type of subsample will however be in Section 4 when we
examine clustering. It will be useful to us to also look at halos ranked
by mass for a fixed number of subhalos. In this case we will use the same
technique, breaking the sample into a large number of narrow bins,
but this time in subhalo number.

\subsection{Spatial distribution}

As an example of our
categorization of halos by occupation
we show in Figure \ref{slice} the spatial distribution of high
and low occupation halos. Here high occupation halos are the top 5\%
by number of subhalos per unit mass
at fixed mass, using the definition given above
(Section 2.3), and the low
occupation are the bottom 5 \%. For reference, the mean number of
subhalos in the high occupation sample is 17.6 and the mean mass
per halo is $3.8 \times 10^{12} \msun$. The mean number of subhalos for
the bottom 5\% sample is 4.1 and the mean mass per halo is $3.9 \times
10^{12} \msun$. 

 Looking at Figure \ref{slice} we can immediately see that there is 
a difference in the way the two subsamples are tracing out structures.
Lower occupation halos appear to outnumber the high occupation halos in 
the low density regions that fill most of space. In the densest areas,
there are many more high occupation halos. We will see in the following
Sections 3.1 and 3.3 that this is borne out quantitatively by considering
the halo occupation distribution in different density environments
and also measuring the correlation function for the two samples we are
plotting here.

\section{Halo Occupation Distribution}
Many observational measurements of galaxy clustering can be reproduced by
theories that include two main ingredients, the clustering of dark
matter halos, and a model for the number of galaxies in a halo.
The Halo Occupation Distribution (HOD, see e.g. Berlind \& Weinberg 2002, 
Zheng \etal 2002, Kravtsov \etal 2004) is a way of formulating the latter.
In the case were galaxies
are identified with dark matter subhalos, the HOD can be measured from 
a simulation by simply counting the number of subhalos as a function
of halo mass. In the present paper we will only concern ourselves
with the mean number of galaxies in a halo, leaving the
form of the probability distribution for further work. We note that 
Boylan-Kolchin \etal (2010) have found that
subhalo abundances are not well described by Poisson statistics at low mass,
but rather are dominated by intrinsic scatter.

\begin{figure*}
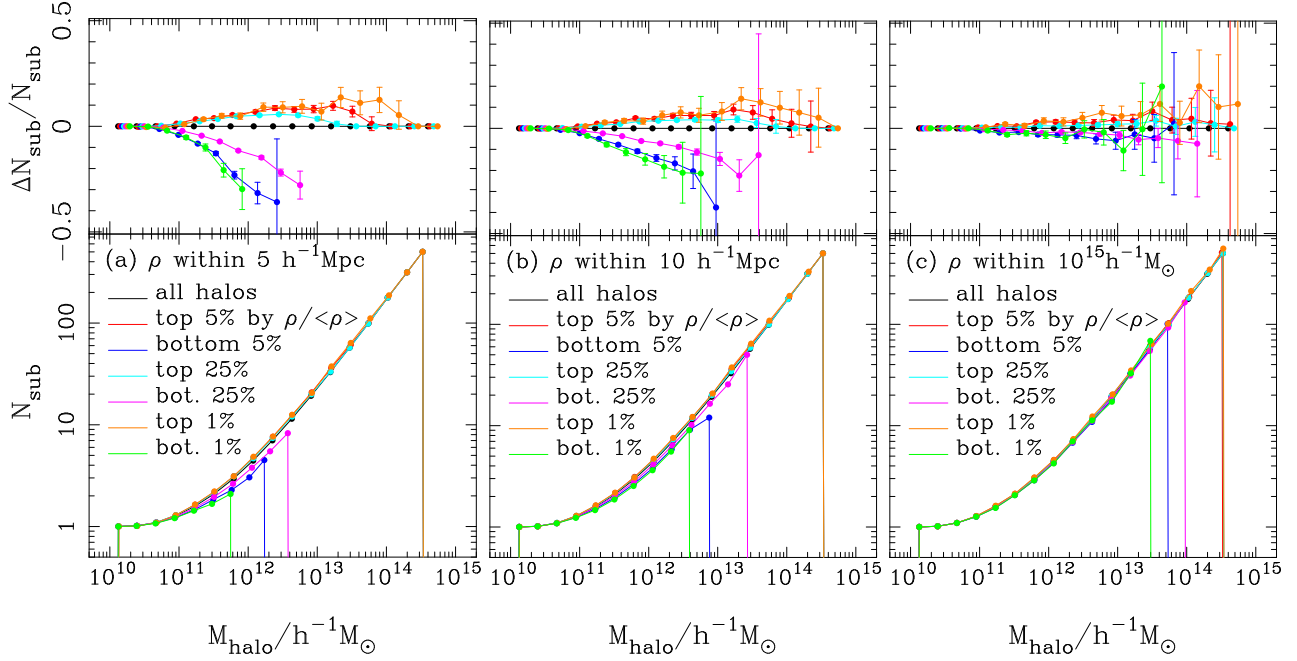

\centerline{
\psfig{file=hodrho1.ps,angle=-90.,width=6.3truecm}
\psfig{file=hodrho2.ps,angle=-90.,width=5.22truecm}
\psfig{file=hodrho3.ps,angle=-90.,width=5.22truecm}
}
\caption{
The halo occupation distribution as a function of environment. The black
lines
in the bottom panels show the HOD for all halos. The red 
lines show
the HOD for the top $5\%$ of halos ranked by the local density and 
the other colours for other percentiles, as listed in the legend.
 The local density is given by averaging
in a sphere of radius $5 \hmpc$ around each halo (left), $10 \hmpc$ (middle),
or the sphere that contains a mass of $10^{15} \msun$ (right). The top panels
show the fractional differences from the HOD for all halos.
The mass limit for a subhalo in all cases was $6 \times 10^{9} \msun$.
We show jackknife error bars on the points (see text).
\label{hod}
}
\end{figure*}

The simplest assumption that can be made about the HOD is that
the mean number of subhalos, $N_{\rm sub}$ depends only on halo mass. This 
assumption has been shown to hold relatively well by Berlind \etal (2003) in
(hydrodynamic) simulations and is consistent with the e.g.,
Yang \etal  (2005) observational
measurements of galaxy groups. It is known however that substructure
fraction does depend on environments in simulations (e.g., Wang \etal 2011).
We will see
how this translates into changes in the HOD, first by looking at
the local overdensity around halos and then some other
measures of the environment.

\subsection{Density dependence}
We measure the density within a radius of 
each halo center of mass. We then rank
the halos by density and plot the HOD for the halos as a function
of halo mass for different density subsamples. Results are shown in panels 
(a) and (b) of Figure \ref{hod}, where we show the HOD for the 
top 5\% by density, bottom 5\% by density and for the whole sample. In
the top panels of that figure we show the fractional difference from
the HOD for the whole sample for the two extreme density bins.
To compute the error bars on the fractional difference, 
the volume was split into octants, and
a jacknife estimator (Bradley 1982) was used to compute the error on the mean
from the standard deviation of the jacknife subsamples.

We can see from Figure \ref{hod}(a) that the number of subhalos in a halo
does depend on the local density, with the halos located in the densest
($5\%$) of environments having a peak difference in 
halo occupation of $\sim10 \%$ more subhalos than the set of
all halos. This difference is
most prominent for halos of masses $10^{12} -10^{13} \msun$ and becomes
zero at lower and higher halo mass. The difference is even larger for
halos in underdense regions, which have less subhalos than the set of all
halos by up to $\sim 40\%$, a result which again depends on halo mass.
Looking at more extreme ends of the $N_{\rm sub}$ distribution, the peak
shifts to the right (e.g., $\sim10^{13.5}\msun$ for the top $1\%$ by
occupation.

The environmental dependence  continues out to larger radius, as can be
see in Figure \ref{hod}(b) where we use density measured with $10 \hmpc$
to rank halos. A third method to measure the density is that within 
a Lagrangian volume with mass $10^{15} \msun$. This is shown in panel (c),
where the sign of the effect is the same, although the amplitude for
the low density environments is very different (and therefore the
effect is detected at a lower level of
significance). This is likely to do with the fact that the
lowest density environments in this Lagrangian picture are being 
measured out to very large radii and therefore diluting the effect.
For example, the mean radius that encloses $10^{15} \msun$
for the ``bottom 1\%'' line is $23 \hmpc$.

The dependence of halo properties on environment has been investigated by
many authors (e.g., Ishiyama \etal 2008, Wang \etal 2011). Recently 
Jeeson-Daniel \etal (2011) have shown
that environment does not  correlate with substructure mass fraction
on a halo by halo basis.
This appears to be at odds with what we find here, but
the definition of environment used by Jeeson-Daniel \etal (2011)
is different to ours, being chosen so that it is not dependent
on halo mass. Also we count subhalos by number and not
by mass fraction. Wang \etal (2011) on the other hand
have found a relationship between local tidal field of
a halo and substructure. 
Jeeson-Daniel \etal and Skibba \etal (2011) find that concentration
(closely related to age) is more fundamental in setting a
range of halo properties than mass. 

The destruction of 
substructure over time in halos can explain the anticorrelation
between age or concentration and substructure (Gao \etal 2004). One might
therefore expect there to be less substructure in halos which formed
earlier, and have a higher concentration. Our results could be explained
therefore if halos in higher density regions had later formation times
and lower concentrations, but this is not the case in general (as pointed out
by W06 and GW07), so that the situation is more complex.
 In W06 and Wetzel \etal (2007) it was shown that low concentration halos 
and late
forming halos do preferentially reside in high density environments,
but only provided the halo masses are $M>M_{*}$. Zentner (2007) pointed out
that  is the general dependence that would be expected from an
excursion set theory analysis (see also further work by Dalal \etal 2008).
These papers argued that high-concentration
and early-forming halos are found in high density environments when 
$M < M_{*}$
because these halos have their growth quenched by the tidal fields of nearby
large halos. At $z=1$, $M_{*} \sim 10^{11}\msun$ so this effect is more relevant
than at $z=0$ where almost all halos of interest have $M>M_{*}$.

\begin{figure*}
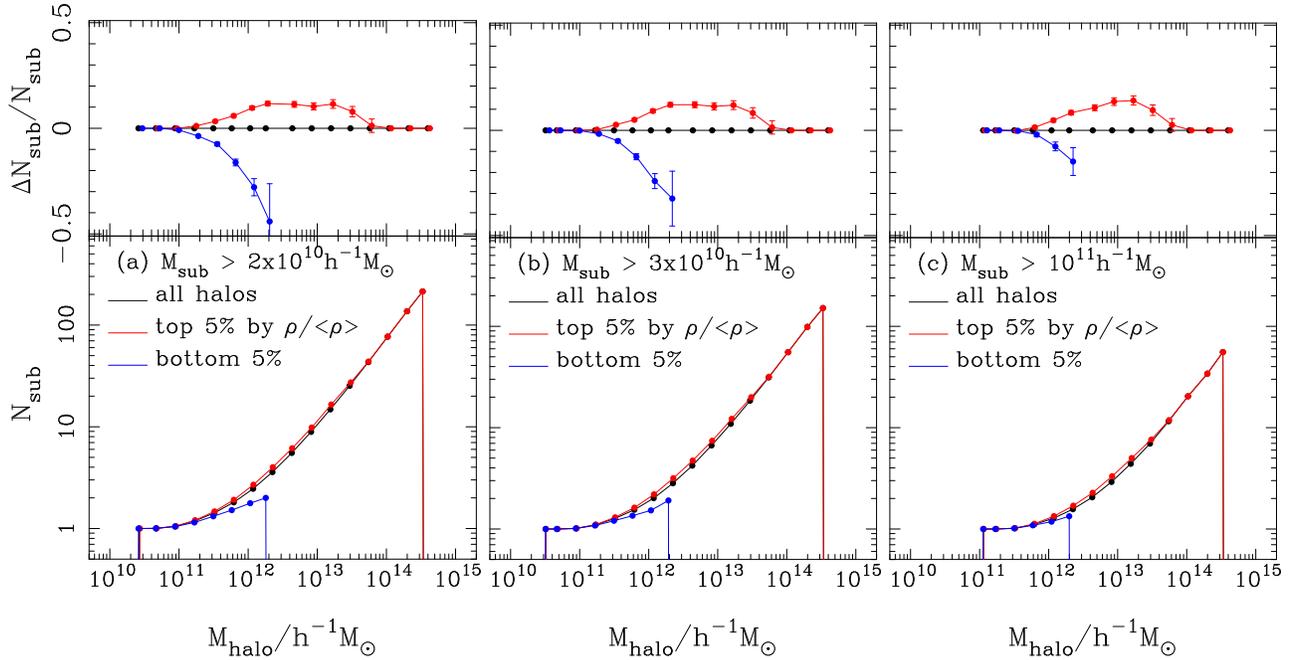

\centerline{
\psfig{file=hodrho64.ps,angle=-90.,width=6.3truecm}
\psfig{file=hodrho95.ps,angle=-90.,width=5.22truecm}
\psfig{file=hodrho320.ps,angle=-90.,width=5.22truecm}
}
\caption{
The halo occupation distribution as a function of environment for 
different lower limits on the subhalo mass (panels from 
left to right). As in Figure \ref{hod} 
the black lines
in the bottom panels show the HOD for all halos and the red and blue lines show
the HOD for the top $5\%$ of halos ranked by the local density and 
bottom $5\%$, respectively. The local density is given by averaging
in a sphere of radius $5 \hmpc$ around each halo. 
The top panels
show the fractional differences from the HOD for all halos.
We show jackknife error bars on the points (see text).
\label{hod2}
}
\end{figure*}

  Gioccoli \etal (2010) have plotted the HOD for halos identified at
different redshifts, finding a systematic shift over all masses
to more subhalos at higher redshifts (and higher 
concentration), which is consistent with the
interpretation of our finding above
(see also Kravtsov \etal 2004, and Zentner \etal 2005).
We do however only find a difference
in HOD over a limited range of masses, so that other effects are involved
as well.
 
One thing which is clear is that the halo model of galaxy clustering
which assumes a fixed HOD for all enviroments can be wrong at up
to the $40\%$ level in the least dense environments. How this dependence
of HOD on environment affects statistics that are measured, such
as the correlation function of galaxies is best addressed by computing them
directly, which we do in Section 4.

Our minimum mass to be counted as a subhalo is $6\times 10^9\msun$ 
(20 particles).  In Figure \ref{hod2} we show the effect on the environmental
dependence (defined as density within 5 $\hmpc$)
 of the HOD on changes in this parameter.
In panels (a)-(c) we are varying the subhalo mass threshold by  factors
of 3,4 and 16. The number of subhalos at fixed mass obviously decreases
sharply as the  subhalo mass threshold is raised. We however do not
see any noticeable change in the difference between subhalo number in
low and high density environments. This is interesting as one may
have thought that small mass halos would be more susceptible to
destruction and so there should be more of an effect.
This also does not appear to be a resolution effect (we return to this
below).

In both Figure \ref{hod} and Figure \ref{hod2} we can see that the
environmental dependence of the HOD appears to be much smaller, or go away 
completely for very high mass host halos ($m_{\rm halo}\simgt 10^{13} \msun$).
This appears to be in accord with Zentner \etal (2005) and W06, who found that 
for low occupation number halos the environmental trends are those induced
by formation time (because for small halos the abundance of subhalos 
reveals a lot about the mass accretion history of the
host halo.) For larger halos this relationship is not as strong and so one
might expect the environmental trend to be less evident also.

\subsection{Resolution test}

\begin{figure}
\centerline{
\psfig{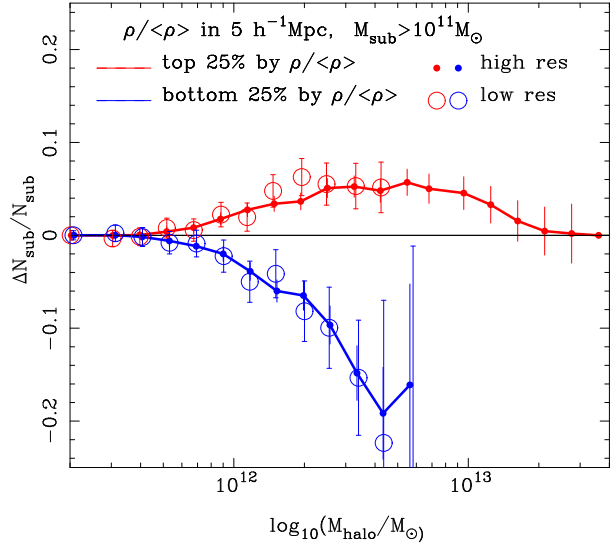}}
\caption{
\label{restest}
A resolution test of the effect of local environment on the HOD.
We show the fractional difference between the number of subhalos in 
all halos and in the top $25 \%$ picked by local density (red)
and bottom  $25 \%$ picked by local density (blue) (similar to 
top panels of Figure \ref{hod}). Results for our fiducial simulation
are shown as smalled filled points and results for a simulation with 
8 times worse mass resolution and 
smaller box size (see Section 3.2) are shown as open
circles.
}
\end{figure}

As a test of the effect of resolution on our results, we have run a simulation
with an identical cosmology but at worse mass resolution, and in a
smaller volume. The simulation had  $512^{3}$ particles in a box
of side length $167 \hmpc$, so that the mass per particle is 8 times larger
than in our fiducial simulation (the volume is also 14 times smaller).
We again identified subhalos using {\small SUBFIND},
and compared the HOD in both 
simulations where the subhalo mass threshold was 40 particles in the low
resolution run and 320 in the fiducial simulation (the same mass in each).

 The results are shown in 
Figure \ref{restest}, where we show the 
fractional difference between mean number of
subhalos for all halos and the top and bottom $25\%$ selected by mass,
(similar as in the top panels of Figures \ref{hod} and \ref{hod2}
except here we show only the top and bottom quartiles because of the small
number of halos in the low resolution simulation). 
We can see the pattern again, with more subhalos in denser regions 
and a rapid dropoff at high
masses in the number of subhalos in underdense regions. There are not
enough halos with masses above $\sim 5 \times 10^{12} \msun$ in the smaller,
low resolution run to be able to compare results, but below that mass there
does not appear to be any systematic difference between the $N_{\rm sub}$ values
and those for our standard simulation.
If numerical effects were responsible for the destruction of subhalos in 
high concentration regions one would expect there to be significantly
different amounts of substructure.
 Given that this does not occur despite the large difference in mass
resolution  is some evidence for the robustness of our results.

\subsection{Other environmental factors}

Although local overdensity is often used interchangeably with environment,
other local measures of the environment of halos have been shown to affect
halo properties. For example, Wang \etal (2011) find that at fixed halo
mass, halo properties depend strongly on the local tidal field. 
The substructure mass fraction in their simulations is particularly
affected, although the nature of the dependence is complex. They suggest
that halos in
higher density (and higher tidal field regions) have more
accretion and so more substructure.

\begin{figure*}
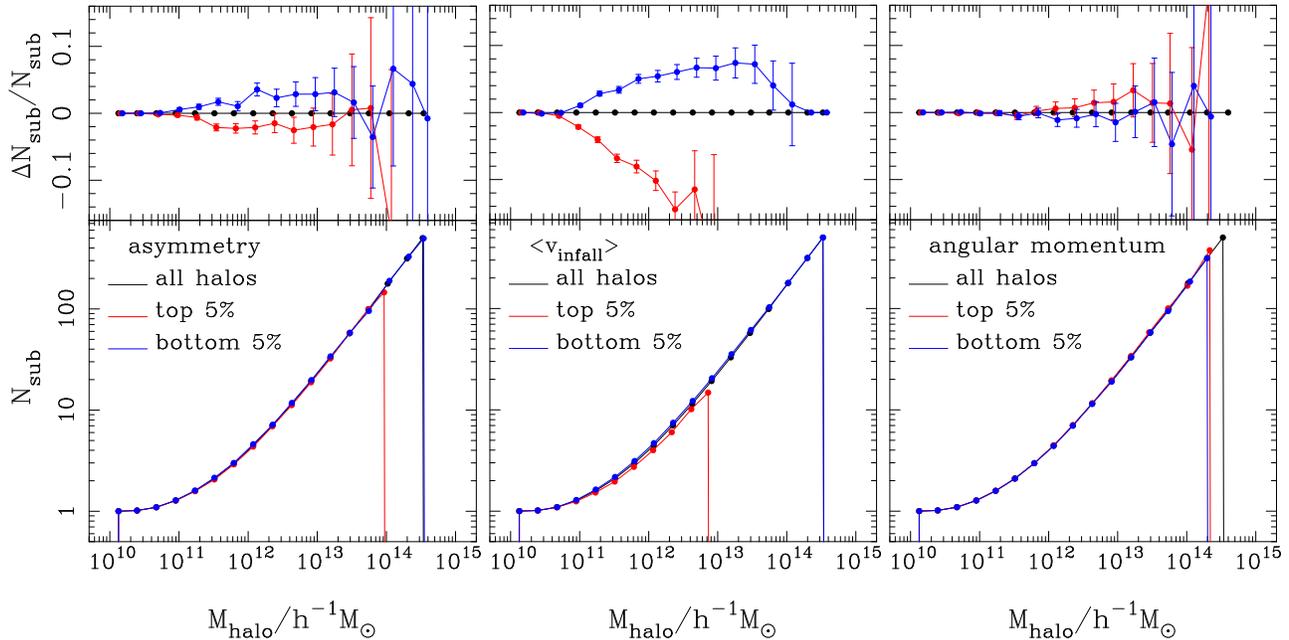

\centerline{
\psfig{file=hodasymm.ps,angle=-90.,width=6.3truecm}
\psfig{file=hodvinfall.ps,angle=-90.,width=5.22truecm}
\psfig{file=hodangmom.ps,angle=-90.,width=5.22truecm}
}
\caption{
The halo occupation distribution as a function of environment for 
different measures of environment (all measured from matter in a $5 \hmpc$ 
sphere
surrounding the halo). The panels from 
left to right show respectively the asymmetry of the matter distribution
(quantified by the difference between the center of mass of the 
matter with $5 \hmpc$ and the center of the halo), the mean infall velocity
of matter, and the total angular momentum of matter. As in Figure \ref{hod} 
the black lines
in the bottom panels show the HOD for all halos and the red and blue lines show
the HOD for the top $5\%$ of halos ranked by the environmental
measure in question and for  
bottom $5\%$, respectively. 
The top panels
show the fractional differences from the HOD for all halos.
\label{hod3}
}
\end{figure*}

We investigate three measures of local environment apart from the density
and look at their effect on the HOD. Again we use a fixed radius of $5 \hmpc$
to define the local region around each halo. We look at the asymmetry of
the local mass distribution (defined as the offset of the center of mass
 and the center of the halo, divided by $5 \hmpc$), the mean infall 
velocity of particles, and the mean angular momentum.
 Of these, the asymmetry has some
similarities with the tidal field studied by Wang \etal (2011), and
the infall velocity will be strongly related to the overdensity.
 The results are shown in 
Figure \ref{hod3}, where we again plot the HOD for the top and bottom
$5\%$ of halos chosen according to each statistic. Again our
fiducial cutoff in subhalo mass is $6 \times 10^{9} \msun$.

We compute the mean (mass-weighted) infall velocity by summing the
mass times the component of the velocity of each particle in the radial
direction (towards the center of the halo) and then dividing by
the total mass of particles within the ($5 \hmpc$) radius.   
As expected, the high mean infall velocity selected halos 
(middle panel) show the same
sign of effect in overabundance of subhalos as seen in highly overdense regions.
The size of the positive effect is again close to a maximum 
$10\%$ difference from
all halos. The negative effect, in halos of low infall velocity is about
a maximum of
$15 \%$ below the result for all halos, somewhat less than the equivalent
result for density (left panel of Figure \ref{hod}.
The asymmetry results are more instructive, with the most asymmetric
local regions having fewer subhalos (by $\sim 2\%$) than all halos, and
the least asymmetric $\sim 2\%$ more. That this is a separate effect to 
the density effect which can be understood by  the fact
that the most asymmetric regions actually
have higher densities (because they have more neighbouring halos).
Looking at the relationship between density and asymmetry
(not plotted) we find for the mean asymmetry,
 $A\simeq 0.15+0.25\log\rho_{5 \hmpc}$
where $\rho_{5 \hmpc}$ is the density within $5 \hmpc$. We also find that the
mean value of the density $\rho_{5 \hmpc}$, for the halos with the top $5\%$
of asymmetry $A$ values is the same as for the top $7\%$ of
halos ranked by density $\rho_{5 \hmpc}$. If increased local
density is 
causing the trend, one should therefore expect (based on Figure 2) to 
find a $10\%$ increase in the number of subhalos in the most asymmetric
regions, rather than a $2\%$ decrease. 

There appears to be little dependence of halo substructure on
angular momentum of the surrounding region (3rd panel), although
the error bars are large.

\section{Clustering}

Given that the environment on at least 5 $\hmpc$ and 10 $\hmpc$ scales
noticeably affects the number of subhalos in a halo at fixed mass, 
a natural progression is a study of the clustering properties of halos.
As described in Section 2.3 we have constructed samples of halos with
varying numbers of subhalos and fixed halo mass, as well as samples
of fixed mean halo occupation but varying mass. In this section
we will examine the clustering of these various
samples. Previous work on substructure
and clustering includes that of W06, who computed
the mark correlation function of halos, with $N_{\rm sub}$ as the mark,
 and G07 who used substructure mass fraction as a defining variable for
samples of halos..

\subsection{Autocorrelation function}

\begin{figure*}
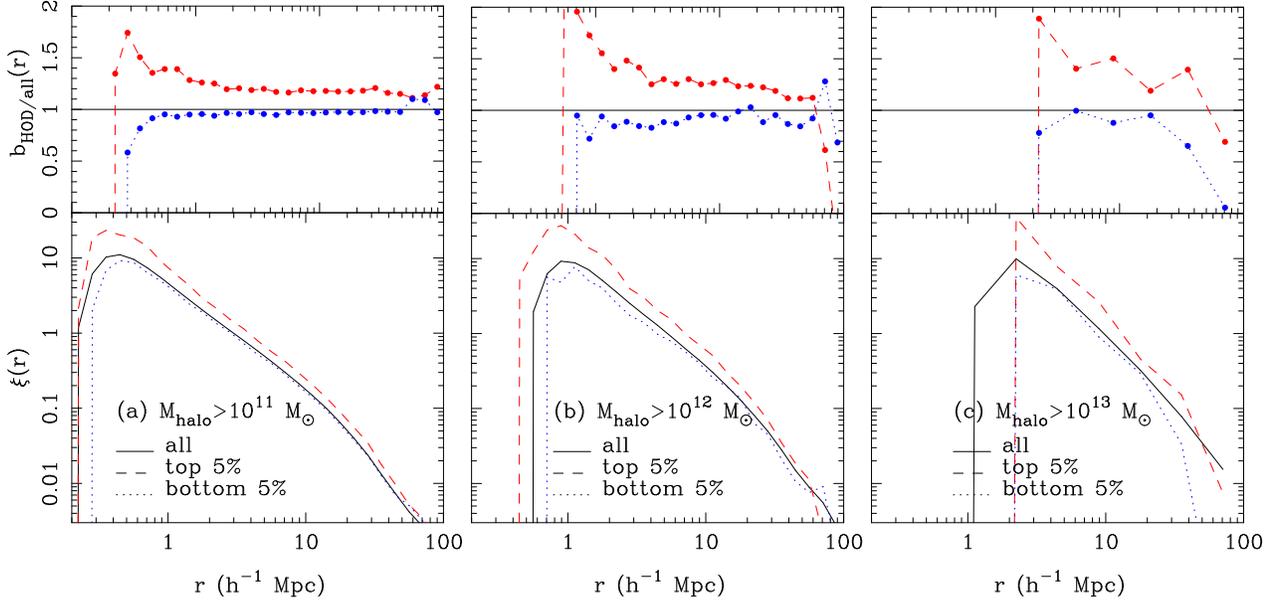

\centerline{
\psfig{file=xibias1.ps,angle=-90.,width=6.08truecm}
\psfig{file=xibias2.ps,angle=-90.,width=5.22truecm}
\psfig{file=xibias3.ps,angle=-90.,width=5.22truecm}
}
\caption{
\label{xibias}
The autocorrelation function  $\xi(r)$
of dark matter halos at fixed mass (Section 2.3) for different
halo occupations. We show different lower mass limits on the halo
mass in the panels from left to right.
In each lower frame,  $\xi(r)$ for all halos above
the mass limit is shown as a solid line and dashed and dotted lines are
used to denote  $\xi(r)$ for the top and bottom $5\%$ of halos ranked
by occupation. The top frames show the relative bias $b_{\rm HOD/all}(r)$ obtained by
dividing the curves for the top and bottom $5\%$ of halos ranked by occupation
by the curve for all halos.
}
\end{figure*}

In Figure \ref{xibias}, we show the autocorrelation function $\xi(r)$ 
of halos for 
three different mass thresholds, as a function
of separation, $r$. In each case, we have selected samples
of fixed mean halo mass but with occupation in the top $5\%$ of halos,
and bottom $5\%$ of halos. The results for the middle panel (lower mass
limit of $10^{12} \msun$) therefore correspond to the halos plotted in 
Figure \ref{slice}. We can see that in each panel the autocorrelation
function of the high occupation halos is significantly enhanced with respect
to that of all halos (also shown). The low occupation halos, on the other
hand have a slightly lower amplitude of clustering. 

We examine this relative bias in the top panels of Figure
\ref{xibias}, where we plot the relative bias with respect to all halos,
as a function of scale, e.g. for high
occupation halos,

\begin{equation}
b_{\rm HOD/all}(r)=\sqrt{\xi_{\rm top 5\%}(r)/\xi_{\rm all}(r)}.
\end{equation}

We can see in Figure \ref{xibias} that the relative bias is close to flat
as a function of $r$ for low occupation halos, and for high occupation it is 
flat on large scales, $r > 5 \hmpc$ for all mass bins. The effect of
high occupation is more pronounced for halos of larger mass, with the
$> 10^{13} \msun$ halos having a relative bias $b_{\rm HOD/all} \sim 1.4$ 
on large scales (an amplitude
of $\xi \sim$ twice that of all halos, and the low mass cutoff halos $m >
10^{11}\msun$ having $b_{\rm HOD/all} \sim 1.2$ on these scales.

Just as we expected from the plot of halo positions, there is a substantial
clustering difference between halos of different occupation at
fixed mass. How exactly this bias is related to halo occupation and
mass can be examined by working with measurements of the 
large scale (flat part) of the bias directly, which we do below.

\subsection{Large-scale bias vs. mass and occupation}

\begin{figure}
\centerline{
\psfig{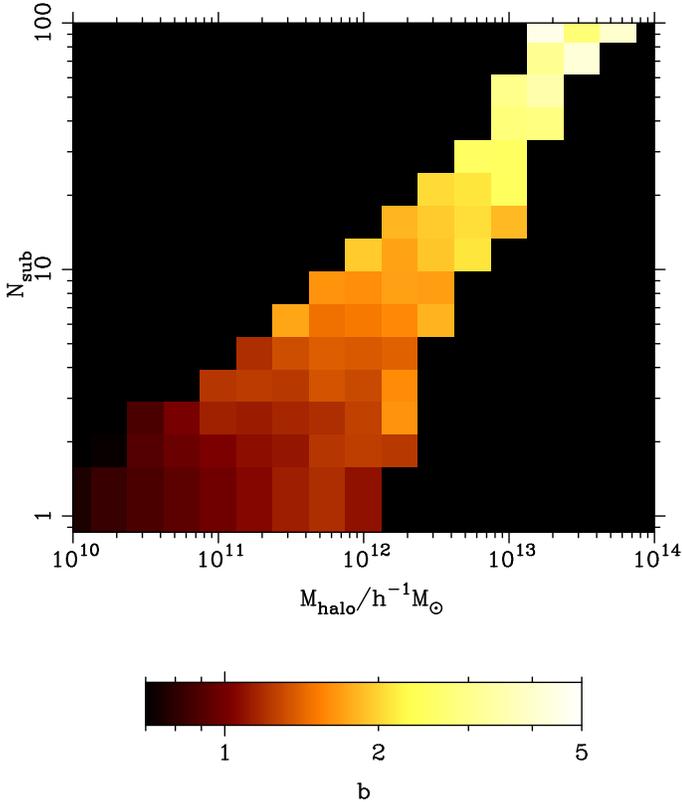}
}
\caption{
\label{biasgrid}
Large-scale bias of halos with respect to the dark matter
as a function of halo mass and halo occupation. The
autocorrelation function $\xi(r)$ was computed for subsamples of halos defined
by $M_{\rm halos}$ and $N_{\rm sub}$. The large-scale (a fit to points
with $r> 5 \hmpc$) bias $b$ for these
subsamples was computed 
with respect to the autocorrelation of dark matter particles (see Section
4.1) and
is shown as a colour scale on the plot. Regions which are black are areas
of parameter space ($M_{\rm halos}$, $N_{\rm sub}$) for
which there are no halos.
}
\end{figure}

Because a focus of this paper is
on the link between clustering and halo occupation and mass we compute
bias for a grid of values of these two parameters. To measure
the large-scale bias $b$ between halos and dark matter we use $\xi(r)$
data points with $r> 5 \hmpc$. We describe
our jacknife estimator for computing the error on $b$ below.
 In Figure
\ref{biasgrid} we show results for $b$ 
on a colour scale, as a function of halo mass
and subhalo number along the two axes. 
The plane of the figure is not filled, but the relatively large scatter
about the mean relation between $N_{\rm sub}$ and halo mass means that there
is data for a substantial fraction of parameter space that lies off this
relation. For example, at low mass and low  $N_{\rm sub}$ there are
$10^{12} \msun$ halos with only 1 subhalo of mass greater
than our threshold ($6 \times 10^{9} \msun$), as well as some halos of similar
mass but with $N_{\rm sub}$ a factor of 15 larger. 

The most obvious trend in the $b$ values shown in the plot is the increase 
in $b$ along the mean relation between $N_{\rm sub}$ and $M_{\rm halo}$. Halos
of larger mass tend to cluster more and also on average tend to have more
subhalos. It is also clear at low masses and low subhalo numbers that 
increasing halo mass but keeping $N_{\rm sub}$ fixed (moving along rows from 
left to right) yields an increasing $b$, as does increasing 
 $N_{\rm sub}$ but keeping  $M_{\rm halo}$ fixed (moving up columns). It is
interesting that from this it appears that $N_{\rm sub}$  and $M_{\rm halo}$
independently control the amplitude of clustering. This holds in the bottom
left of the plot, but when one moves to the top right a trend along rows
or up columns is harder to see. By using an averaging technique (Section 4.3
see below)
we will see that it turns out that at high occupation, $b$ is
completely independent of $M_{\rm halo}$, (no trend 
along rows in Figure \ref{biasgrid}) whereas at high masses, $b$ is
still correlated with $N_{\rm sub}$ (still a trend up columns, at least
for the mass range shown).

\begin{figure}
\centerline{
\psfig{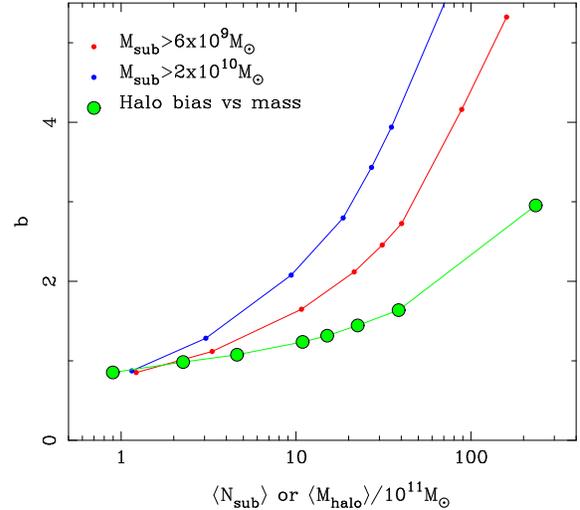}
}
\caption{
\label{biassubandmass}
Large scale bias between halos and dark matter as a function of
either halo mass (large points) or halo occupation
(small points). For each point a lower limit was applied
to the sample of all halos and the position of the point on the $x$-axis
corresponds to the mean value (of either $M_{\rm halo}$ or $N_{\rm sub}$) of
halos above the cut. For $N_{\rm sub}$ we show two curves for two different
lower mass cuts on the subhalo mass.
}
\end{figure}

Before examining these trends in $b$ as a function of halo occupation 
at fixed mass, we will explore the variation of $b$ with halo mass
and $N_{\rm sub}$. The steepness of these trends will allow us to
place our later results in context. 
That halo bias is strongly
influenced by halo mass is a result that is at the heart of many analyses
of large scale stucture (e.g., Kaiser 1984, 
Mo \& White 1996, Seljak \& Warren 2004)
In Figure 
\ref{biassubandmass} we show $b$ (in this case
the bias of halos with respect to the dark matter distribution)
plotted against mean halo mass in 
various mass bins, finding qualitatively the same steep relationship
seen in e.g., Figure 8 of   Seljak and Warren (2004).

 We also plot 
$b$ against the mean number of subhalos in a halo, finding a steeper 
relationship (in that $b$ increases faster for a given increase
in $\log N_{\rm sub}$). Because $N_{\rm sub}$ increases more slowly than 
$M_{\rm halo}$ as $M_{\rm halo}$ is increased (see e.g., Figure \ref{hod}) this
behaviour is expected. We show $b$ as a function of $N_{\rm sub}$
for two different subhalo mass thresholds, finding that $b$ increases
faster for a higher threshold mass. This again is expected as for higher
mass thresholds the HOD curves stay longer on the shallower initial
part of the $N_{\rm sub}-M_{\rm halo}$ relationship (see e.g., Figure \ref{hod2}). 

\subsection{Large scale bias at fixed halo mass and at fixed occupation}

\begin{figure}
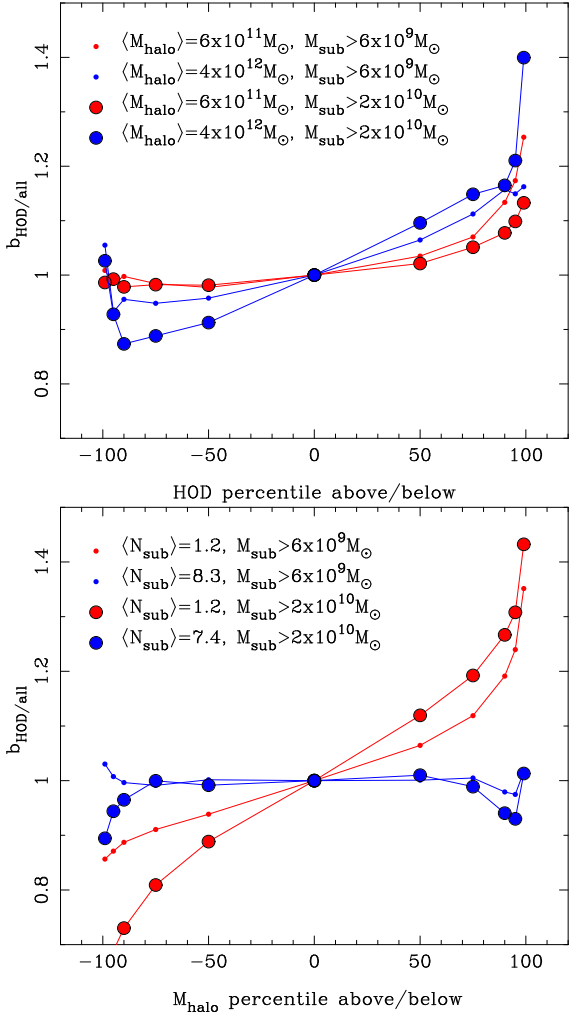

\centerline{
\psfig{file=bias1.ps,angle=-90.,width=7.5truecm}
}
\centerline{
\psfig{file=biasfixednsub1.ps,angle=-90.,width=7.5truecm}
}
\caption{
\label{bias1}
Large-scale bias $b_{\rm HOD/all}$ 
as a function of halo occupation percentile for 
fixed halo mass (top
panel) and halo mass percentile for fixed halo occupation (bottom panel).
 The definition
of the percentile scale is given in Section 4.3.
The $b_{\rm HOD/all}$ 
values shown are the relative bias between the halos and the
subsample of halos in the percentile of either halo occupation of mass.
All curves therefore pass through $b_{\rm HOD/all}=1$ at $x=0$ (see Section
4.3). 
In the top (bottom) panel we show results for two values of mean halo
mass (occupation). For each of these we show what happens when the 
subhalo mass limit is changed (from $6\times10^{9}\msun$ to 
$2\times10^{10}\msun$).
}
\end{figure}

We take samples of halos which are chosen to
either have fixed halo mass or fixed number of subhalos (as defined
in Section 2.3), and compute how the large scale bias
$b_{\rm HOD/all}$ changes as we vary the other parameter. Our results are shown in 
Figure \ref{bias1}, where on the $x-$axis we show the percentiles
of the distribution. For example in the top panel a point at $x=+99\%$
shows $b_{\rm HOD/all}$ for the top $1\%$ of halos by occupation, at fixed halo
mass. Likewise a point at $x=-99\%$ shows $b_{\rm HOD/all}$ for the bottom 
$1\%$ of halos by occupation, at  $x=-50\%$ for the bottom $50\%$
and so on. Points at $x=0$ show results for all halos in that halo
mass bin, and the bias $b_{\rm HOD/all}$ 
is measured relative to this (as in equation 1).
This is different from $b$ computed with respect to the dark matter
distribution, and ensures that all curves pass through $b_{\rm HOD/all}=1$
at $x=0$.

In the top panel of Figure \ref{bias1}, the values of  $b_{\rm HOD/all}$
 are shown for 
variations in the $N_{\rm sub}$ percentile, at fixed halo mass. We show
results for two different halo masses, and two different cutoffs in
the mass of a subhalos. The mean halo mass $\langle M_{\rm halo}\rangle$
shown in the figure
caption for the curves was computed by averaging over
the masses of halos above a mass threshold. As required, 
the $\langle M_{\rm halo}\rangle$ are
approximately
constant for the different bins of  $N_{\rm sub}$ percentile, and 
in all cases within $10\%$ of the $\langle M_{\rm halo}\rangle$ value given.
  We can see that  $b_{\rm HOD/all}$ does change significantly 
with $N_{\rm sub}$ percentile, even though the mean halo mass is the same
for all points along the curve. This is what we expect from looking at Figures
\ref{slice} and \ref{xibias}. We can also see that the steepest change 
in $b$ occurs for the larger mass halos. This analysis
is equivalent to adding together the trend in $b$ that one gets by moving up
different columns in Figure \ref{biasgrid}. 

If we now turn to  the results showing how  $b$ varies as $M_{\rm halo}$ 
is changed for fixed $N_{\rm sub}$ (bottom panel of Figure \ref{bias1}), we can
see that the situation is somewhat different. In this plot,  the 
$\langle N_{\rm sub}\rangle$ values shown are 
computed from the average number of
subhalos above a threshold in  $N_{\rm sub}$, for each of the bins of
$M_{\rm halo}$ percentile. Here we see that for a low number of subhalos,
 $\langle N_{\rm sub}\rangle=1.2$, there is a strong trend of $b$ with halo mass,
one which is strong for both values of subhalo mass cutoff.  However when
we increase the $N_{\rm sub}$ threshold to give a higher value
of $\langle N_{\rm sub}\rangle\sim 8$, we 
find that there is no dependence of  $b_{\rm HOD/all}$
on $M_{\rm halo}$ percentile.

\begin{figure}
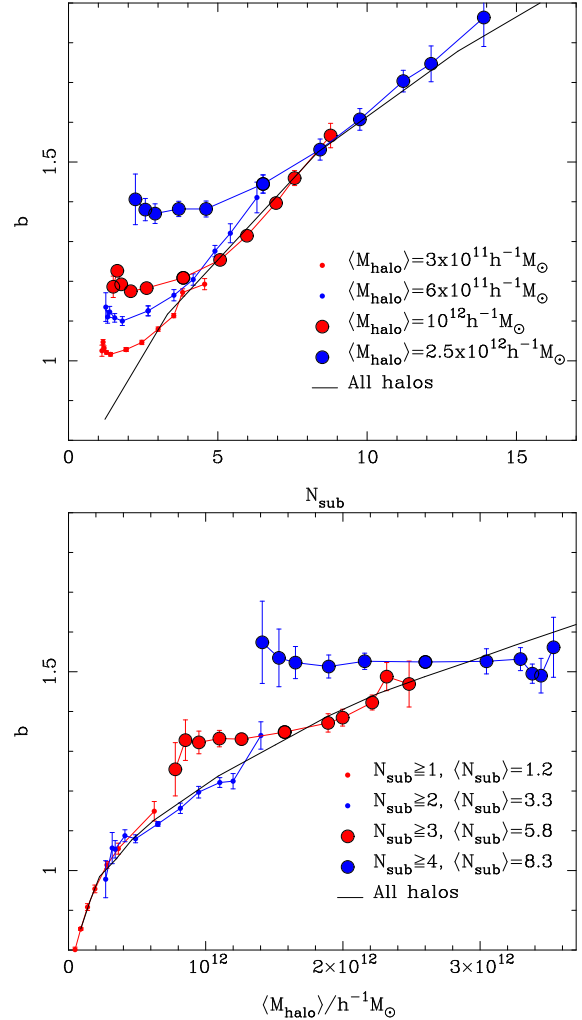

\centerline{
\psfig{file=bias2.ps,angle=-90.,width=7.5truecm}
}
\centerline{
\psfig{file=bias3.ps,angle=-90.,width=7.5truecm}
}
\caption{
\label{bias2}
Large-scale bias (wrt. dark matter) for halos of fixed mean
mass but varying $N_{\rm sub}$ (top panel) and halos of fixed  $N_{\rm sub}$
but varying $M_{\rm halo}$ (bottom panel). In the top panel for reference
 we show $b$ vs.  $N_{\rm sub}$ for all halos as a solid black line. Each
set of other lines in the top panel is for a fixed mean halo mass (given
in the legend) and shows how $b$ varies with $N_{\rm sub}$  in that case.
In the bottom panel, $b$ vs. $M_{\rm halo}$ is shown as a solid black line.
Each of the other lines in the bottom 
panel is for a fixed mean  $N_{\rm sub}$ (given
in the legend) and shows how $b$ varies with $M_{\rm halo}$.
In all cases we apply the usual lower mass limit on the subhalo mass of
$6 \times 10^9 \msun$.}
\end{figure}

This independence of halo clustering and
halo mass is interesting enough that we revisit it in Figure \ref{bias2}.
where we now turn to plotting $b$ with respect to 
dark matter halo clustering. The scale of the effects can then be judged
compared to the overall trend of clustering increasing with greater
halo mass and with greater $N_{\rm sub}$. 

 In order to judge the 
significance of the comparison, it is useful to put error bars on the points.
We have done this by first splitting the simulation into octants
and using a jackknife estimator. Using random
catalogs we have computed $\xi(r)$ for the halo subsamples and for
the dark matter particles, for the simulation volume minus each octant in
turn. The value of $b$ was computed for each jackknife subsample, with
the error bar coming from their standard deviation.

In Figure \ref{bias2}, as in Figure \ref{bias1}, the top panel shows
$b$ as a function of $N_{\rm sub}$ for fixed halo mass, and the bottom panel
show $b$ as a function of $M_{\rm halo}$ for fixed $N_{\rm sub}$.  We show
results for 4 different halo masses in the top panel, and the $x-$axis
is $N_{\rm sub}$ instead of $N_{\rm sub}$ percentile. We can see that even
for  $M_{\rm halo}=4\times10^{12}\msun$ there is a trend of increasing $b$
with increasing  $N_{\rm sub}$. As in  Figure \ref{bias1}, 
for the smallest $N_{\rm sub}$ 
we notice that there is some flattening and even a small increase in $b$.
For larger $N_{\rm sub}$ the curves for varying $N_{\rm sub}$ at fixed 
halo mass appear to track the black line relatively well, which shows 
how $b$ varies with  $N_{\rm sub}$ for all halos.

\begin{figure}
\centerline{
\psfig{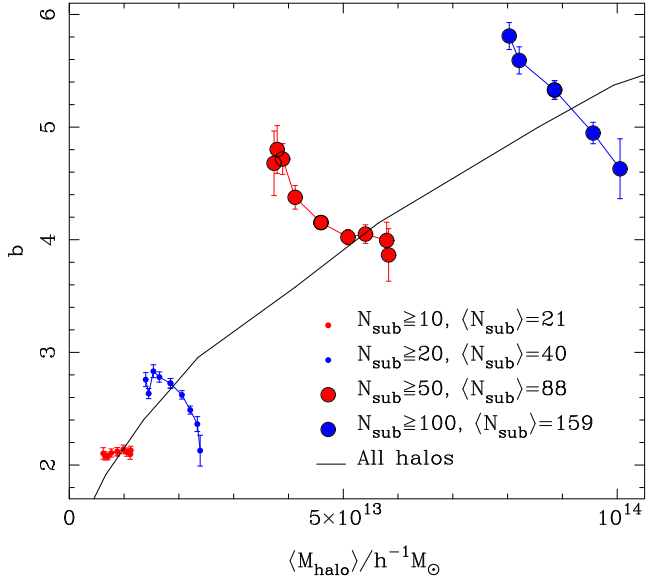}
}
\caption{
\label{largehalos}
Large-scale bias (wrt. dark matter) for halos of fixed mean   $N_{\rm sub}$
but varying $M_{\rm halo}$.
This plot is identical in nature to the bottom panel of Figure \ref{bias2}, but
shows results for larger halos. 
}
\end{figure}

In the bottom panel, the results for fixed
$N_{\rm sub}$ show a different behaviour. 
We see that for a sample with $N_{\rm sub}$ threshold of 4 subhalos 
(with mass $> 6\times10^{9} \msun$ ) there is no dependence of $b$ on halo 
mass. To show how different this behaviour is from that usually seen
in the halo mass-bias relation (e.g., Seljak \& Warren 2004)
we can compare to the black line in the bottom panel of Figure \ref{bias2}
which shows how $b$ 
varies as a function of halo mass for all halos.  The 
relative change in $b$ for all halos is substantial (a change of $35\%$)
over the mass range that is probed by the blue line (where there is
no change in $b$). 

The behaviour of this relationship is examined for larger halos in Figure
\ref{largehalos}. We show the same bias versus mass plot as in
the bottom panel of Figure \ref{bias2}, but this time for halos with a 
lower limit on $N_{\rm sub}$ varying from 10 to 100. We can see that for
fixed subhalo number, the  lack of dependence of bias on halo mass 
seen in Figure \ref{bias2} gradually turns into a strong
anticorrelation of bias on halo mass. This happens 
between a threshold of  $N_{\rm sub}=10$ and $N_{\rm sub}=20$, halos which have
masses of $\sim 10^{13} \msun$ and above.

 We have seen earlier (Figure
\ref{hod}) that the environmental dependence of the HOD declines
for halos larger than $\sim 10^{13} \msun$. From our clustering results ( Figure
\ref{largehalos})
one might perhaps have expected a negative environmental dependence which is
not seen in Figure \ref{hod}. The situation is undoubtedly complex.

\section{Summary and Discussion}

\subsection{Summary}
Using a large, high resolution dark matter simulation 
output at redshift $z=1$, we have 
investigated the role that the environment of halos plays on the
number of subhalos they host, and we examined the relationship between
clustering and subhalo number. We find that

(1) At fixed halo mass, the number of subhalos in a halo is 
affected by local density, with overdense regions having 
more substructures over the whole range of halo masses, and underdense regions
having less. This effect can be as large as $40 \%$ for the most
underdense $5\%$ of regions.

(2) Our finding (1) is not significantly affected by the mass limit applied
to subhalos, or (in a resolution test) by simulation resolution.

(3) At a much smaller level, the asymmetry of the local density (quantified
by a centroid shift) affects the subhalo number, so that the most asymmetric
halos have fewer subhalos, at the percent level.

(4) The clustering of halos at fixed mass is strongly affected by the
number of subhalos. This is true over the entire mass range tested, from 
$10^{11}-10^{13} \msun$ and for different values of subhalo mass cutoff.
As with prior examples of the dependence on clustering of
variables other than mass (e.g. age, Gao \etal 2005,
concentration, W06), it is thus straightforward to generate samples 
of halos which have the same mass but
widely different clustering properties. 

(5) The clustering of halos at fixed number of subhalos is only positively
dependent on mass for small halos. As we increase the size of halos,
we find that for halos with more than 4 subhalos (for 
a subhalo mass limit of $6\times10^{9}\msun$), at fixed
number of subhalos there is no dependence of clustering strength on halo mass,
and then for larger halos, of group and cluster size, we find that this
turns into a strong anticorrelation of clustering amplitude and halo mass.
This case, showing  first an independence of clustering and halo mass 
and then an anticorrelation of the two is perhaps 
the bluntest  expression yet of the shortcomings of the standard halo model
when dealing with the subtleties of galaxy clustering.

\subsection{Discussion}

The most widely used
version of halo model of galaxy clustering has at its heart some very simple
premises. 
It is already well known however from simulation studies
that many parameters apart from halo
mass are actually at 
work in the translation of halo clustering into galaxy clustering,
including the formation time dependence of clustering
(Gao \etal 2005), substructure
dependence  and concentration dependence (W06). Looking for weak spots
in the predictions of the halo model can in principle make the search
for better models easier. It can also highlight areas in which for the
present non-linear simulations can be a necessary tool in the attempt to 
make precise predictions of galaxy clustering. The areas we have investigated
in this paper may have some observational consequences (as we discuss below),
but they also serve to highlight some special cases for which the
halo model's prime assumptions are in direct contradiction with what is seen.
These include a sample of halos of the same mass which have radically
different clustering properties based on an internal property (subhalo number)
as well as other samples of halos with very different masses but the 
same clustering amplitude.

The environmental dependence of the HOD is a relatively subtle
effect, as can be seen by the fact that some
previous
smaller  simulations (e.g.  Berlind \etal 2003) were consistent with 
no dependence. Also,
the observed correlation function of galaxies can be well modelled
by an environmentally independent HOD (Zehavi \etal 2004, Zheng \etal 2009). 
There are however already signs that some observed galaxy statistics
are not well predicted by the halo model, including higher
order clustering of SDSS galaxies in low density environments 
(Berrier \etal 2011).

As future galaxy surveys move well past the million redshift regime
(e.g., Schlegel \etal 2011), we can ask however
how well we need to know the HOD and any extra 
environmentally dependent terms in order to carry out
precision cosmology with an HOD approach to galaxy clustering
(e.g., Zheng \etal 2002). Further work is needed to determine the effect on for 
example the correlation function of including an environmental 
term in the HOD. We have also not investigated in this paper the 
effect of environment on the distribution of the number of halos at
fixed mass (e.g., Kravtsov \etal 2004) which may be strongly affected.

It is also not clear how relevant and how strong the effects that we
have seen here with dark matter simulations are on the galaxies that 
form within the subhalos. One can clearly enumerate many possible
environmental effects that rely on baryonic physics (e.g.,
luminosities of backsplash galaxies, Pimbblet 2011, 
etc..). which will further complicate the effect of 
environment on the HOD, and which may even have the opposite sign.

The obvious example where the number of subhalos is used to
define a set of objects are optical cluster catalogs. The 
environmental dependence of halo occupation is likely 
responsible for a  fraction
of the scatter in the mass-richness relation noted in optical 
cluster finders (see e.g. Rozo \etal 2011 for 
different sources of scatter). 

The cause of the HOD enviromental effect is a complex issue. It has
been shown by many authors that a wide range of variables are affected 
by halo environment such as concentation (e.g. Wang \etal 2011).
internal halo properties such as substructure and shape are
nearly all correlated (Jeeson-Daniel \etal 2011, Skibba
\& Maccio 2011) and the mass function of subhalos 
itself correlates with halo concentration, formation time etc (Gao \etal
2011). The relative importance of mergers and smooth accretion in building
up the mass of halo (Wang 2011) 
is likely to play a role in the enviromental dependence
of the HOD. Fakhouri and Ma (2010) have shown that mergers dominate halo growth
in overdense regions and diffuse accretion in voids. If they survive
the merger and accretion process 
we therefore expect subhalos to be more numerous in overdense regions
(for halos of a given mass), as we have found (see also Wetzel \etal 2007).
 The destruction of
subhalos over time by intrahalo merging and stripping will make this
relationship even more diffcult to decipher.

Incorporating environmental dependences into the halo model, making 
it more complex, but able to deal with phenomena such as those that have 
been demonstrated here is one avenue which can be pursued. Recently, 
Gil-Marin \etal (2011) have presented some first steps in this direction.

Another source of potential uncertainty in the predictions of the halo
model is the definition of halo mass.  More \etal (2011)
have shown that the mass of a FOF halo in a simulation depends on 
resolution. More \etal state that the influence of substructures
(which depend on redshift and 
cosmology) on
the FOF halo boundary,
will make it difficult to model this effect in general. 
In our work, the relationship between the number of substructures
and environment is likely also influenced by the effect
that substructures have on the mass definition.

Direct computation of the dependence of the correlation function of 
halos for different occupations shows more of the complex relationship
between halo properties and clustering. The bias of halos of the same
mass can vary widely depending on their occupation. For example 
(from Figure \ref{bias2}), the lowest $25\%$ of halos by occupation 
at fixed mass can have a bias which is $25 \%$ lower than that for all halos.
This is similar to the effect seen by GW07 based on substructure fraction
within  the FOF group (although not within $r_{200}$) and appears to be a
stronger trend than that based on other properties at fixed mass, such as 
formation redshift (GW05), concentration (W06) or spin (GW07). 

In this paper we have focused on the clustering of halos of
galaxy mass, and also only looked at $z=1$. Further work is needed to
explore the  relationship between clustering and halo occupation in 
cluster size halos and those at lower redshift. If the same
relationships hold, then this could have interesting
consequences for the clustering of galaxy clusters selected in optical
surveys. One could make measurements of the dependence of galaxy cluster bias
on mass (measured using velocity dispersion or lensing mass) and richness,
equivalent to halo occupation. Mapping out the bivariate distribution of
bias values as in Figure \ref{biasgrid} would give further clues to
the nature of galaxy formation in groups and 
clusters and how much it is affected by the non-baryonic processes 
investigated here.

Our final perhaps suprising finding is that one can easily
select samples of halos (by picking a fixed occupation) 
for which there is either no dependence of clustering on
mass or even a strong anticorrelation between the two.
This finding is one which could also be tested 
with observational data on both mass and occupation. It
again points to the complexity of halo clustering and the difficulties
of using galaxy clustering measurements for precision cosmology.

\section*{Acknowledgments}
This work was supported by NSF Award OCI-0749212 and the Moore Foundation.
The research was supported by allocation of advanced computing
resources provided by the National Science Foundation. Simulations
were performed on Kraken at the National Institute for Computational
Sciences (http://www.nics.tennessee.edu) and analysis on facilities
provided by the Moore Foundation in the McWilliams Center for 
Cosmology at Carnegie Mellon University.
RACC would like to thank Michael Busha, Neal Dalal, Alexie Leauthaud, 
Jeremy Tinker, David Weinberg and Andrew Zentner for
useful discussions. We also thank Andrew Zenter for suggesting changes 
which were incorporated into the 
manuscript.

{}

\end{document}